\documentclass[aps,prl,twocolumn]{revtex4-1}
\usepackage{mathtools}
\usepackage{amsmath,amssymb}
\usepackage{graphicx}

\usepackage[usenames]{color}
\usepackage[bookmarks=true,colorlinks,linkcolor=red,urlcolor=blue,citecolor=blue]{hyperref}
\usepackage{natbib}
\input epsf
\usepackage{epstopdf}

\definecolor{darkgreen}{rgb}{0,0.4,0}
\definecolor{darkred}{rgb}{0.4,0,0}
\definecolor{darkblue}{rgb}{0,0,0.4}

\input epsf
\newcommand{\vev}[1]{\langle #1 \rangle}

\newlength{\extraspace}
\newlength{\extraspaces}
\def\ep{\epsilon}

\def\p{\partial}

\def\tr{{\text{tr}}}
\def\Re{{\text{Re}\hskip0.1em}}
\def\Im{{\text{Im}\hskip0.1em}}

\def\vev#1{\langle{#1}\rangle}

\def\II{\relax{I\kern-.10em I}}

\def\IZ{\relax{\rm Z\kern-.34em Z}}
\def\IB{\relax{\rm I\kern-.18em B}}
\def\IC{{\relax\hbox{$\inbar\kern-.3em{\rm C}$}}}
\def\ID{\relax{\rm I\kern-.18em D}}
\def\IE{\relax{\rm I\kern-.18em E}}
\def\IF{\relax{\rm I\kern-.18em F}}
\def\IG{\relax\hbox{$\inbar\kern-.3em{\rm G}$}}
\def\IGa{\relax\hbox{${\rm I}\kern-.18em\Gamma$}}
\def\IH{\relax{\rm I\kern-.18em H}}
\def\II{\relax{\rm I\kern-.18em I}}
\def\IK{\relax{\rm I\kern-.18em K}}
\def\IP{\relax{\rm I\kern-.18em P}}

\def\inbar{\,\vrule height1.5ex width.4pt depth0pt}

\def\p{\partial}

\def\IR{\relax{\rm I\kern-.18em R}}
\def\pbar{\bar{\p}}

\def\lp10{\ell_p^{10}}
\def\lp11{\ell_p^{11}}
\def\R11{R_{11}}

\def\frac#1#2{{#1 \over #2}}

\def\ahat{\hat{A}}

\def\pd#1#2{\frac{\partial #1}{\partial #2}}

\newdimen\tableauside\tableauside=1.0ex
\newdimen\tableaurule\tableaurule=0.4pt
\newdimen\tableaustep
\def\phantomhrule#1{\hbox{\vbox to0pt{\hrule height\tableaurule width#1\vss}}}
\def\phantomvrule#1{\vbox{\hbox to0pt{\vrule width\tableaurule height#1\hss}}}
\def\sqr{\vbox{%
  \phantomhrule\tableaustep
  \hbox{\phantomvrule\tableaustep\kern\tableaustep\phantomvrule\tableaustep}%
  \hbox{\vbox{\phantomhrule\tableauside}\kern-\tableaurule}}}
\def\squares#1{\hbox{\count0=#1\noindent\loop\sqr
  \advance\count0 by-1 \ifnum\count0>0\repeat}}
\def\tableau#1{\vcenter{\offinterlineskip
  \tableaustep=\tableauside\advance\tableaustep by-\tableaurule
  \kern\normallineskip\hbox
    {\kern\normallineskip\vbox
      {\gettableau#1 0 }%
     \kern\normallineskip\kern\tableaurule}%
  \kern\normallineskip\kern\tableaurule}}
\def\gettableau#1 {\ifnum#1=0\let\next=\null\else
  \squares{#1}\let\next=\gettableau\fi\next}

 \def\eqnn#1{\xdef #1{(\secsym\the\meqno)}\writedef{#1\leftbracket#1}%
 \global\advance\meqno by1\wrlabeL#1}
 \def\eqna#1{\xdef #1##1{\hbox{$(\secsym\the\meqno##1)$}}
 \writedef{#1\numbersign1\leftbracket#1{\numbersign1}}%
 \global\advance\meqno by1\wrlabeL{#1$\{\}$}}
 \def\eqn#1#2{\xdef #1{(\secsym\the\meqno)}\writedef{#1\leftbracket#1}%
 \global\advance\meqno by1$$#2\eqno#1\eqlabeL#1$$}

\global\newcount\itemno \global\itemno=0

\def\itemaut#1{\global\advance\itemno by1\noindent\item{\the\itemno.}#1}

\def\({\left(}
\def\){\right)}

\newcommand{\bigvev}[1]{\left\langle #1 \right\rangle}
\newcommand{\zbar}{\bar{z}}
\newcommand{\re}{\Re}
\newcommand{\im}{\Im}

\def\boundaryMat{K} 

\usepackage[yyyymmdd,hhmmss]{datetime}
\newif{\ifeq}           
\eqtrue                 
\usepackage{float}
\usepackage{cleveref}
\newcommand{\Gc}{\mathcal{G}}
\newcommand{\Jc}{\mathcal{J}}

\def\ahat{\hat{a}}
\def\dhat{\hat{d}}
\def\chat{\hat{c}}
\def\cidx{m}
\def\didx{n}
\begin{document}
\title{Eigenvalue spectra of large correlated random matrices} 
\author{Alexander Kuczala and Tatyana O. Sharpee}
\affiliation{Computational Neurobiology Laboratory, Salk Institute for
  Biological Studies, La Jolla, California, USA}
\affiliation{Department of Physics, University of California, San
  Diego, USA}
\begin{abstract}
  Using the diagrammatic method, we derive a set of self-consistent
  equations that describe eigenvalue distributions of large correlated
  asymmetric random matrices. The matrix elements can have different
  variances and be correlated with each other. The analytical results
  are confirmed by numerical simulations. The results have
  implications for the dynamics of neural and other biological
  networks where plasticity induces correlations in the connection
  strengths within the network. We find that the presence of
  correlations can have a major impact on network stability.
\end{abstract}

\maketitle

Random matrices serve as a useful tool for analyzing the stability and
dynamics of a variety of networks, from neuroscience
\cite{sompolinsky1988,Rajan2006,aljadeff2015transition,aljadeff2015low} and genetic circuits \cite{Aldana2007} to
ecology \cite{may1972will,allesina2015}. Spectra of random matrices also help determine solutions
to problems in nuclear \cite{wigner1958distribution} and condensed matter physics
\cite{anderson1958absence,sompolinsky1982} as well as in data compression
\cite{majumdar2009large,candes2006near}.
In particular, the rightmost eigenvalue (the eigenvalue with largest real component) determines the stability of the system's linear dynamics and onset of chaos of the nonlinear dynamics.
Knowledge of the onset of chaos is also useful for determining the network's computational capabilities \cite{Sussillo2009,Bertschinger2004} as well as  the network's response to inputs \cite{Rajan2010}.

However, most of these results do not address an important feature of biological circuits where
connection strengths are correlated
\cite{Gilson2009,Song2005,Miner_2016}. While correlated Hermitian ensembles have received some attention, \cite{vinayak2010,shukla2005random,khorunzhii1996eigenvalue,burda2005spectral}, results about correlated non-Hermitian ensembles are scarce \cite{sommers1988spectrum,rogers2016modularity}.
Most notably, the correlations in the connection strengths arise as
the result of plasticity, where connections are modified depending on
node activity and network input.  One of the predominant effects of
plasticity is that it induces correlations between forward and reverse
connections \cite{Gilson2009,Miner_2016}. That is, the degree to which
node $i$ affects node $j$ is correlated with the strength of the
reverse connection from node $j$ to node $i$. We focus here on this
circuit motif when considering correlations between matrix elements.


Consider a network with $N$ nodes $i=1,\ldots N$, with linear dynamics
\begin{equation}
\dot{x_i}(t) = -x_i(t) + \sum_{j=1}^N J_{ij} x_j(t),
\label{eqn:linearSystem}
\end{equation}
where $x_i(t)$ describes the activity of each node and $J$ is the
$N\times N$ connectivity matrix. The solution of this system is
$\mathbf{x}(t) = e^{(\mathbf{1-J})t}\mathbf{x}(0)$. This system has
stable equilibria only if the rightmost eigenvalue of $J$ is less than
one. For networks with nonlinear dynamics, mean-field methods can be
used to show that the transition to chaotic behavior still occurs when
the rightmost eigenvalue of $J$ is $<1$
\cite{sompolinsky1988,aljadeff2015transition,aljadeff2015low}.

In this work we use the diagrammatic approach to analyze the case
where the matrix elements of $J$ are correlated and not identically
distributed.
%
Specifically, we consider an $N\times N$ complex non-Hermitian
Gaussian random matrix $J$ whose elements are distributed according to
\begin{equation}
P(J) \propto \exp\left[-\frac{N}{2}\sum_{i,j} \begin{pmatrix} J_{ij}^* & J_{ji}^* \end{pmatrix} \mathbf{V}^{-1}\binom{J_{ij}}{J_{ji}}\right]
\label{eqn:Jdistribution}
\end{equation}
where covariance matrix $\mathbf{V}$ consists of real-valued variances
\begin{equation}
\vev{J_{ij}J^*_{ij}} = \frac{1}{N} g_{ij}^2,
\label{eqn:variances}
\end{equation}
and real-valued covariances
\begin{equation}
\vev{J_{ij}J_{ji}} = \frac{1}{N} \tau_{ij} g_{ij}g_{ji}.
\label{eqn:covariances}
\end{equation}
All other-second order correlations vanish. The gain matrix $g_{ij}$
has positive elements. Correlation values $\tau_{ij}$ are symmetric in
$i,j$, $|\tau_{ij}|\leq 1$, and denote the degree of correlation
between forward $j,i$ and reverse $i,j$ connections in the
corresponding random network.

To outline the steps of the derivation, we will first seek the
expected density of eigenvalues of $J$ for large $N$ by first writing
the density in terms of the Green's function $G$. While $G$ is
analytic for Hermitian matrices, $G$ is generally non-analytic for
non-Hermitian matrices, so we cannot directly apply the diagrammatic
method.
We therefore relate $G$ to the analytic Green's function of a
Hermitian random matrix $H$, which we compute with standard
diagrammatic techniques. We derive a set of self-consistent equations
for $G$ for the case where the gain matrix $g_{ij}$ is a continuous
function in the limit $N\rightarrow \infty$, and the case where
$g_{ij}$ is block-structured. Finally, we apply our method to two
example problems and compare the results to empirical eigenvalue
distributions obtained by exact diagonalization of realizations of
$J$.

\begin{figure}

%
  \includegraphics[width=3.5in]{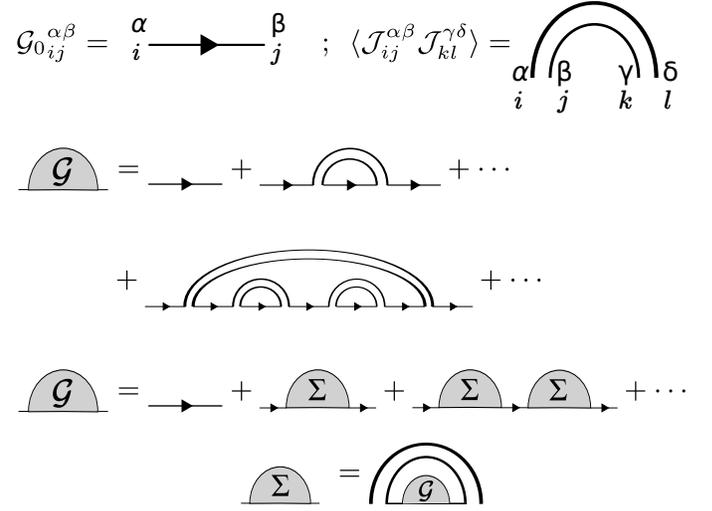}
\caption{Diagrams used in the expansion (\ref{eqn:propExpand}) of $\Gc$.
  $\Gc$ is the sum of all planar diagrams in the large
  $N$ limit. $\Gc$ can be re-summed in terms of the self-energy matrix
  $\Sigma$. In the large $N$ limit, $\Sigma$ consists of all diagrams
  nested under a double line (\ref{eqn:sigma}).}
\label{fig:feynmanRules}
\end{figure}

We start by writing the expected density of eigenvalues of $J$
in the complex plane as
\begin{equation}
\rho(x,y) = \bigvev{\frac{1}{N}\sum_k \delta(x - \re \lambda_k)\delta(y-\im \lambda_k)}.
\label{eqn:density}
\end{equation}
where $\vev{\cdot}$ indicates an average over realizations of $J$ according to Eq. (\ref{eqn:Jdistribution}).
Defining $\p = (\p_x-i\p_y)/2$ and $\pbar =
(\p_x+i\p_y)/2$, and using the identity $\pbar \frac{1}{x+ iy} = \pi
\delta(x)\delta(y)$
\footnote{This relation follows from the solution $\pbar \p
\log z = \pi \delta(x)\delta(y)$ of Poisson's equation in two dimensions. See also \cite{zee1997}.},
we can write the density (\ref{eqn:density}) in terms of the Green's function 
\begin{equation}
G(z,\zbar)\equiv\bigvev{\frac{1}{N}\tr \frac{1}{z-J}}
\label{eqn:green}
\end{equation}
as
\begin{equation}
\rho(x,y) = \frac{1}{\pi}\pbar G(z,\zbar).
\label{eqn:rhoAndG}
\end{equation}
Since $J$ is non-Hermitian, the eigenvalues of $J$ will in general lie
in some region of the complex plane. For example, Ginibre's circular law states that
if the elements of $J$ are independently and identically distributed with variances
$g^2/N$, then the eigenvalues lie in a disk of radius $g$ \cite{ginibre1965statistical}.
The Green's function is therefore not in general holomorphic, and we cannot expand in powers of $1/z$ as required for the diagrammatic expansion.
Following \cite{zee1997}, we  can find the Green's function by
solving a related Hermitian random matrix problem, to which we can apply the diagrammatic approach.  Define the $2N \times 2N$
Hermitian matrix
\begin{equation}
H=\begin{bmatrix}
0 & J - z \\ (J - z)^\dagger & 0 \\
\end{bmatrix}.
\end{equation}
The matrix Green's function for $H$ is
\begin{equation}
\Gc(\eta,z,\zbar) = \bigvev{\frac{1}{\eta - H}},
\label{eqn:GcDef}
\end{equation}
where we think of the eigenvalues of $H$ as lying on the complex plane
$\eta$. Since $H$ is Hermitian, these eigenvalues will lie on the real
axis, and $\Gc$ is holomorphic in $\eta$ except for cuts on the real
axis. 
Once $\Gc$ is computed, we obtain the original Green's function $G$ from $\Gc$ 
by extracting the lower left matrix block and taking the limit $\eta \to i0^+$:
\begin{equation}
 \Gc(\eta=0,z,\zbar) = \bigvev{\begin{bmatrix}
 0 &\frac{1}{(z-J)^\dagger}  \\ \frac{1}{z-J} & 0 \\
 \end{bmatrix}},
 \end{equation}
 yielding Eq. (\ref{eqn:green}):
\begin{equation}
	\label{eqn:lowerLeftBlock}
	G(z,\zbar) = \frac{1}{N}\tr \Gc^{21}(\eta=0,z,\zbar).
\end{equation}
Here, $\Gc^{21}$ is the lower left block of $\Gc$.
To compute $\Gc$ (\ref{eqn:GcDef}), we first rewrite
$\eta - H
 = \Gc_0^{-1}-\mathcal{J}$
with
\begin{equation}
\label{eqn:matrixDefs}
\Gc_0^{-1}\equiv\begin{bmatrix}
\eta & z \\ \ \zbar & \eta  \\
\end{bmatrix}
\quad \text{and} \quad
\Jc\equiv \begin{bmatrix}
0 & J \\ J^\dagger & 0  \\
\end{bmatrix},
\end{equation}
so that the random part $\Jc$ has zero mean. Note that $\Gc_0$ is just $\Gc$ with $J=0$.
We expand $\Gc$ in $\Gc_0$ as follows:
\begin{equation}
\label{eqn:propExpand}
\Gc = \sum_{n=0}^\infty \Gc_0\vev{(\Jc \Gc_0)^n} = 
\Gc_0 + \vev{\Gc_0 \Jc \Gc_0 \Jc \Gc_0} + \ldots
\end{equation}
Here, the odd terms vanish since $\vev{\Jc}=0$. Since the distribution over $\Jc$ is Gaussian, each term in the sum reduces to the Wick contraction of $n$ factors of $\Jc$. We therefore use the
diagrammatic technique \cite{brezin1978planar,brezin1994correlation} to represent each term in the sum.  We
denote the $N$ node indices by roman letters $i=1,\ldots N$ and
index the blocks by Greek letters $\alpha=1,2$. 
We represent $\Gc_0$ by a single directed line carrying one set of indices, and the correlator $\vev{\Jc \Jc}$ by a double line carrying two sets of indices (Fig. \ref{fig:feynmanRules}) \cite{zee1997,janik1997non, feinberg2006non}. Indices are summed at each connecting vertex.
The $n$th term in $\Gc$ is the sum of all diagrams with $n$ vertices. 
In the large $N$ limit
diagrams which have crossing lines vanish, and only ``planar''
diagrams remain \cite{t1973planar,ahmadian2015}
\footnote{It is worth noting that since the correlators
  (\ref{eqn:variances}, \ref{eqn:covariances}) are not proportional to
  the identity as in the i.i.d. case, loops produce a weighted trace
  weighted by elements of $g_{ij}$. However, assuming all elements of
  $g_{ij}$ are of $O(1)$, the weighted trace is of $O(N)$ as in the
  i.i.d. case.}.  This greatly simplifies the sum, since the only
allowed diagrams are nested `rainbow diagrams' such as those depicted
in Fig. \ref{fig:feynmanRules}.
This allows us to evaluate (\ref{eqn:propExpand}) by performing a
resummation of $\Gc$ in terms of the `self-energy' matrix $\Sigma$:
\begin{equation}
\Gc = \sum_{n=0}^\infty \Gc_0(\Sigma \Gc_0)^n = \left(\frac{1}{\Gc_0^{-1} - \Sigma}\right).
\label{eqn:GcSigma}
\end{equation}
In the planar limit, the self-energy matrix is
\begin{equation}
\Sigma= \vev{\Jc \Gc \Jc},
\label{eqn:sigma}
\end{equation}
encoding the nested `rainbow' structure of the diagrams \cite{brezin1994correlation}. This is depicted diagrammatically in Fig.  \ref{fig:feynmanRules}.

In block form, Eq. (\ref{eqn:GcSigma}) is
\begin{equation} 
  \Gc = 
  \begin{bmatrix}
  A & B \\ C & D \\
  \end{bmatrix}
  =
  \begin{bmatrix}
  \eta - \Sigma^{11} & z - \Sigma^{12} \\ \zbar - \Sigma^{21} &\eta -  \Sigma^{ 22}\\
  \end{bmatrix}^{-1}.
  \label{eqn:blockGc}
\end{equation}
and Eq. (\ref{eqn:sigma}) is
\begin{equation} 
\Sigma =
\begin{bmatrix}
\Sigma^{11} & \Sigma^{12} \\ \Sigma^{21} & \Sigma^{22}
\end{bmatrix}
=\bigvev{ \begin{bmatrix}
J D J^\dag & J C J \\ J^\dag B J^\dag & J^\dag A J
\end{bmatrix}}
\label{eqn:blockSigma} 
\end{equation}
where we have denoted the blocks of $\Gc$ as $A,B,C$ and $D$. Substituting (\ref{eqn:blockSigma}) into (\ref{eqn:blockGc}) will give us self-consistent equations for the blocks of $\Gc$. 

Equations (\ref{eqn:blockGc}) and (\ref{eqn:blockSigma}) describe the
eigenvalue distribution in the general case, with or without
correlations. Before analyzing the impact of correlations on the
eigenvalue distribution, we first check that this result reproduces
previous results obtained in the absence of correlations.  When
elements of $J$ are independently distributed, the covariances
(\ref{eqn:covariances}) vanish.
In this case we find
\footnote{We furthermore demand that $g_{ij}$ converges to a uniformly
  bounded continuous function $g(i/N,j/N)$ on the unit square as $N\to
  \infty$, excepting discontinuities on a set of measure zero (see
  \cite{aljadeff2015low}).}:
\begin{align}
\Sigma^{11}_{il}&= \sum_{j,k}\vev{J_{ij} D_{jk} J^\dag_{kl} } = \frac{1}{N} \delta_{il}\sum_{j} g_{ij}g_{lj} D_{jj},\\
\Sigma^{22}_{il}&=\sum_{j,k}\vev{J^\dag_{ij} A_{jk} J_{kl} } = \frac{1}{N} \delta_{il}\sum_{j} g_{ji}g_{jl} A_{jj},
\end{align}
and $\Sigma^{12} = \Sigma^{21} = 0$.
This means that the matrix $\Sigma$ is diagonal. Then, since each
block on the RHS of Eq.~(\ref{eqn:blockGc}) is diagonal, each block of
$\Gc$ is also diagonal. Inverting the RHS and equating matrix elements
yields
\begin{equation}
A_{ii} = \frac{\eta - \frac{1}{N}\sum_j A_{jj} g_{ji}^2 }{q_i(\eta,|z|)}, \quad
D_{ii} = \frac{\eta - \frac{1}{N}\sum_j g_{ij}^2 D_{jj}}{q_i(\eta,|z|)},
\label{eqn:AD}
\end{equation}\begin{equation}
C_{ii} = \zbar/q_i(\eta,|z|),
\label{eqn:C}
\end{equation}
where
\begin{equation}
q_i(\eta,|z|) = (\eta-\frac{1}{N}  \sum_j A_{jj} g^2_{ji})(\eta - \frac{1}{N}  \sum_j g^2_{ij}D_{jj}) - |z|^2.
\end{equation}
Writing out the blocks of $\Gc$ in Eq. (\ref{eqn:GcDef}),
\begin{equation}
\begin{bmatrix}
A & B \\ C & D \\
\end{bmatrix}
=
\bigvev{
\begin{bmatrix}
\frac{\eta}{\eta^2 - (J-z)(J-z)^\dag} & \frac{J-z}{\eta^2 - (J-z)^\dag(J-z)}  \\
\frac{(J-z)^\dag}{\eta^2 - (J-z)(J-z)^\dag} & \frac{\eta}{\eta^2 - (J-z)^\dag(J-z)}  \\
\end{bmatrix}
},
\label{eqn:blockGcDef}
\end{equation}
and rewriting $\eta = i\ep$, with $\ep>0$, we see that blocks $A$ and
$D$ are positive definite matrices multiplied by $-i$.
We therefore define $a_j \equiv i A_{jj}$ and $d_j \equiv i D_{jj}$,
where $a_i$ and $d_i$ are positive real numbers. We also define $c_j =
C_{jj}$. This allows us to rewrite (\ref{eqn:AD}) and (\ref{eqn:C}) as
\begin{equation}
a_i = \ahat_i/q_i, \quad d_i = \dhat_i/q_i,\quad c_i  =
\zbar/q_i(\ep,|z|)
\label{eqn:selfConsistentNoCov}
\end{equation}
with $q_i(\ep,r) \equiv -q_i(\eta,|z|) = \ahat_i \dhat_i + r^2$ and
\begin{equation}
\ahat_i \equiv \ep +\frac{1}{N} \sum_j a_{j} g_{ji}^2,\quad
\dhat_i \equiv \ep + \frac{1}{N}\sum_j g_{ij}^2 d_{j},
\label {eqn:ahatdhat}
\end{equation}
where  $r = |z|$.
We now have a set of $2N$ self-consistent equations (\ref{eqn:selfConsistentNoCov}) for the elements $a_i$ and $d_i$ of the Green's function $\Gc$. These can be solved numerically with $\ep=0$ (or
 $\ep$ set to a small value if many elements $g_{ij}$ are also
small). Once the $a_i$ and $d_i$ are found, the $c_i$ can be computed and used to find the 
original Green's function $G$ with  Eq. (\ref{eqn:lowerLeftBlock}), since the trace of $\Gc^{21} \equiv C$ is the sum of the coefficients $c_i \equiv C_{ii}$. Note that since $c_j  = r e^{-i \theta}/q_j(\ep,r)$
in polar coordinates, $|c_j|$ depends only on $r$. This allows us to rewrite Eq. (\ref{eqn:rhoAndG}) as a function of $r$ only:
\begin{equation}
\rho(r) = \frac{1}{2\pi N}\sum_j\left(\pd{|c_j|}{r}+\frac{|c_j|}{r}\right).
\label{eqn:polarDensity}
\end{equation}
The resulting eigenvalue distribution has support on
the disk with radius $r = \sqrt{\lambda_1(\boundaryMat)}$, where $\lambda_1(\boundaryMat)$
is the largest eigenvalue of the matrix $\boundaryMat_{ij} \equiv
g^2_{ij}/N$ (see Appendix).

\paragraph{Symmetric covariances}
We now allow $J$ to have correlated elements across its
diagonal (Eq. \ref{eqn:covariances}).
Then $\Sigma^{12}$ and $\Sigma^{21} \neq 0$, yielding  a new expression for $c$:
\begin{equation}
c_{i} = \chat_i/{q_i(\ep,z,\zbar)}, \quad \chat_i \equiv \zbar - \frac{1}{N} \sum_j \tau_{ij} g_{ij}g_{ji}\bar{c}_{j},
\label{eqn:cCov}
\end{equation}
where now $q_i = \ahat_i \dhat_i + |\chat_i|^2$, $b_i=\bar{c_i}$. The $\tau_{ij}$ denote the degree of correlation between $i$ and $j$ as in Eq. (\ref{eqn:covariances}).
In this case, the eigenvalue density has the more general form
\begin{equation} \rho(x,y) = \frac{1}{\pi}\pbar G(z,\zbar) =
\frac{1}{N\pi}\pbar \sum_{j=1}^N c_j(z,\zbar).
\label{eqn:eigDensityContinuous}
\end{equation}
The density $\rho$ depends on $x$ and $y$ in a nontrivial way, and the support of the distribution is neither circular nor elliptical.
The boundary of the eigenvalue distribution now satisfies (see Appendix for a derivation):
\begin{equation}
\lambda_1(\boundaryMat(z)) = 1,\quad
 \boundaryMat_{ij}(z) = \frac{1}{N}|c_i(z)|^2 g_{ij}^2,
\label{eqn:covBoundary}
\end{equation}
where the complex-valued $c_i(z)$ are now given by the self-consistent equations
\begin{equation}
c_i = (z-\sum_j \tau_{ij}g_{ij}g_{ji} c_j)^{-1}.
\label{eqn:gammaOutside}
\end{equation}
Now, to obtain the boundary, it is necessary to simultaneously solve (\ref{eqn:covBoundary}) and (\ref{eqn:gammaOutside}) for each boundary point. For example, we can set $z = r e^{i\theta}$ and solve the above for $r$ for each $\theta$. Note that these expressions reduce to the circularly symmetric case when $\tau_{ij}=0$.
\paragraph{Block structured}
We now consider the special case for which the gain matrix $g_{ij}$ is
block structured. Block structured matrices describe networks with
nodes partitioned into subgroups, for example neural networks with
cell-type-specific connectivity \cite{aljadeff2015transition}, or
networks of ecological communities \cite{rogers2016modularity}.
Suppose the nodes of the network are grouped into $M$ populations of size $f_\cidx N$, for $\cidx=1\ldots M$ and that $J$ is block structured so that the gain $g_{\cidx_i \didx_j}^2 = g_{\cidx\didx}^2$ and correlations $\tau_{\cidx_i \didx_j} = \tau_{\cidx \didx}$ depend only on the population indices $\cidx$ and $\didx$ of the output and input nodes $i$ and $j$, respectively.
This allows us to sum (\ref{eqn:selfConsistentNoCov}) and (\ref{eqn:cCov}) over each population. 
Let $N_\cidx \equiv N \sum_{\didx=1}^m f_\didx$. Then
define \footnote{These sums converge, since the Green's function for $H$, $\frac{1}{N}\tr \Gc = \sum_i(A_{ii} + D_{ii})$ and Eq. (\ref{eqn:eigDensityContinuous}) must converge}
\begin{equation}
a_\cidx \equiv \frac{1}{N f_\cidx}\sum_{i = N_\cidx+1}^{N_\cidx} a_i,\quad
\end{equation}
and define $c_\cidx$ and $d_\cidx$ similarly.
Then $q_\cidx \equiv q_i$ depends only on the population index, and now we have
\begin{equation}
a_\cidx = \ahat_\cidx/q_\cidx, \quad d_\cidx = \dhat_\cidx/q_\cidx,
\quad c_\cidx = \chat_\cidx/q_\cidx,
\label{eqn:blockEqns}
\end{equation}
and $q_\cidx = \ahat_\cidx \dhat_\cidx + |\chat_\cidx|^2$,
with 
\begin{equation}
\ahat_\cidx = \ep + \sum_{\didx=1}^M f_\didx a_\didx g_{\didx \cidx}^2, \quad
\dhat_\cidx = \ep + \sum_{\didx=1}^M g_{\cidx \didx}^2 f_\didx d_\didx, 
\end{equation} \begin{equation} 
\chat_\cidx = \zbar - \sum_{\didx=1}^M \tau_{\cidx \didx}g_{\cidx \didx}g_{\didx \cidx} f_\didx \bar{c}_\didx.
\end{equation}
Now the dependence on $N$ is removed, and we need only solve $3M$
self-consistent equations.
The eigenvalue density is now 
\begin{equation} \rho(x,y) =
\frac{1}{\pi}\pbar \sum_\cidx f_\cidx c_\cidx(z,\zbar).
\end{equation}
The boundary of the distribution satisfies equations similar to (\ref{eqn:covBoundary}), (\ref{eqn:gammaOutside}), see Appendix.
When $\tau_{\cidx\didx} = 0$, the distribution has boundary $|z| =
\sqrt{\lambda_1(\boundaryMat)}$, where $\lambda_1(K)$ is the largest
eigenvalue of the matrix $\boundaryMat_{\cidx\didx} \equiv g^2_{\cidx\didx} f_\didx$ \cite{aljadeff2015transition}.



\begin{figure}
	\centering
        	\includegraphics[width=3in]{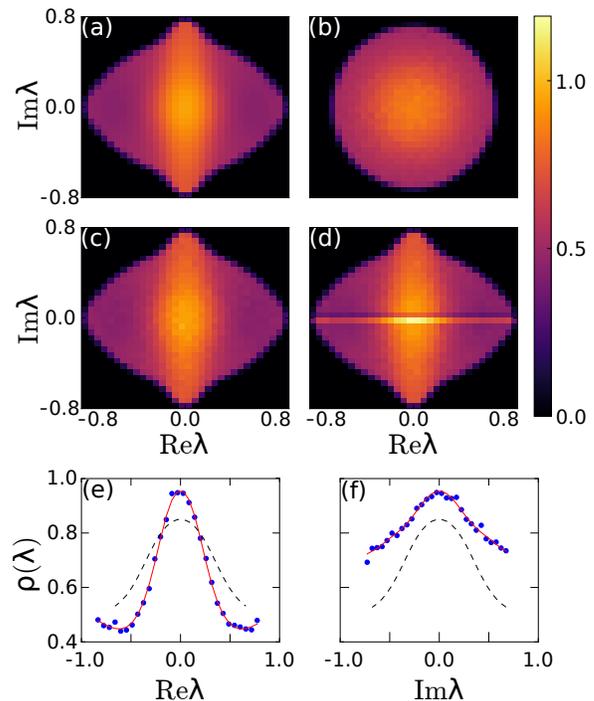}
	\caption{(Color online) Eigenvalue density for block structured $J$ with
          gain and covariance given by (\ref{eqn:blockParams}). (a)
          Eigenvalue density calculated from self-consistent equations (\ref{eqn:blockEqns}).
          (b)
          Empirical histogram of eigenvalues from exact diagonalization of realizations of $J$ with independent elements. The empirical histogram for $J$ with covariance is shown with complex (c) and 
          real (d) entries. (e,f) Cross sections of the density along the
          real (e) and imaginary (f) axes, showing the theoretical result (solid red line), the complex-valued empirical result (blue dots), and the
          distribution with no covariance (dashed curve).}
	\label{fig:blockExample}
\end{figure}

To verify our results, we consider a network with $M=3$ populations, with relative population sizes $f = (1/6,1/3,1/2)$, and
\begin{equation}
g^2_{\cidx\didx} = \begin{bmatrix}
.54 & .83 & .65 \\
.95 & .46 & .01 \\
.72 & .59 & .55
\end{bmatrix},\quad \tau_{\cidx\didx} = \begin{bmatrix*}[r]
.5 & -.2 & .9 \\
-.2 & .3 & .1 \\
.9 & .1 & -.6 \\
\end{bmatrix*}.
\label{eqn:blockParams}
\end{equation}
We iteratively solved the self-consistent Eqs.~(\ref{eqn:blockEqns})
for a grid of points on the complex plane and approximated the
eigenvalue distribution using finite differences, shown in
\crefformat{figure}{Fig.~#2#1{(a)}#3} \cref{fig:blockExample}. We
compare this distribution with eigenvalue histograms generated by
exact diagonalization of 1000 realizations of $J$. We find that
realizations of $J$ with complex elements agree with our result
\crefformat{figure}{Fig.~#2#1{(c,e,f)}#3} (\cref{fig:blockExample}).
Removing the correlations (\ref{eqn:covariances}) from realizations of
$J$ yields a circular distribution
\crefformat{figure}{Fig.~#2#1{(b)}#3}
(\cref{fig:blockExample}). Notably, we find that including these
correlations distorts the eigenvalue distribution in a nontrivial way:
the distribution is neither a circle nor an ellipse. Furthermore, we find
using Eqs. (\ref{eqn:maxEigAppendix}),(\ref{eqn:cOutsideAppendix}) (in the Appendix) that the rightmost
eigenvalue of the distribution has moved from $\sim 0.713$ to $\sim 0.890$, so that
the corresponding linear system (\ref{eqn:linearSystem}) becomes more unstable.

For any finite $N$, $J$ has non-universal features
that disappear as $N\rightarrow \infty$. In particular, the matrix $J$ with real elements will have
a higher density of eigenvalues on the real axis
\crefformat{figure}{Fig.~#2#1{(d)}#3}
(\cref{fig:blockExample}).
However, we find that the proportion of
eigenvalues on the real axis drops off as $1/\sqrt{N}$, as anticipated for large $N$ \cite{edelman1994many}.
\crefformat{figure}{Fig.~#2#1{(e)}#3}
\begin{figure}
	\includegraphics[width=3in]{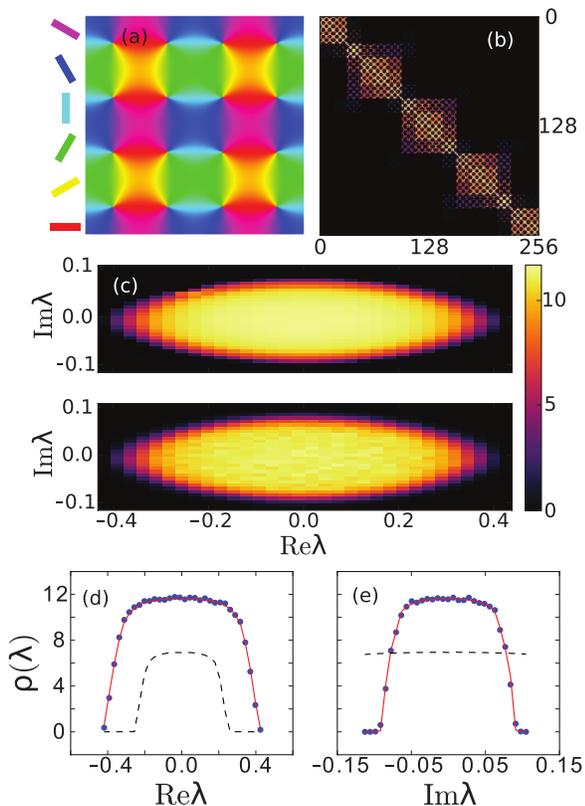}
	\caption{(Color online) Analysis of eigenvalue distribution with continuously
          varying gain (\ref{eqn:continuousGain}). (a) Orientation map of neurons. (b) Gain matrix
          $g_{ij}$. (c) Eigenvalue density calculated from
          self-consistent equations (top) and from realizations of $J$
           (bottom). Density cross sections along the real
          (d) and imaginary (e) axes, plotted as in
          \cref{fig:blockExample}.
          }
	\label{fig:gFunFig}
\end{figure}

To demonstrate that our technique applies to situations where the
variance and covariance depend continuously on the node indices $i,j$,
we consider a neural network inspired by connectivity around pinwheels
in the visual cortex \cite{murphy2009balanced,wolf2005symmetry}. The
neurons are arranged on a square grid on the unit square and assigned
orientations based on their position, shown in
\crefformat{figure}{Fig.~#2#1{(a)}#3} \cref{fig:gFunFig}.  For neurons
$i$ and $j$ with positions $\mathbf{r}_i$ and $\mathbf{r}_j$, the gain
is 
\begin{equation}
g_{ij} =g_0 \exp\left[-|\mathbf{r}_i-\mathbf{r}_j|^2/w_r^2 -
  \Delta\theta^2(\mathbf{r}_i,\mathbf{r}_j)/w_\theta^2 \right],
\label{eqn:continuousGain}
\end{equation}
 where $\Delta\theta(\mathbf{r}_i,\mathbf{r}_j)$ denotes the
 difference in orientation of neurons at $\mathbf{r}_i$ and
 $\mathbf{r}_j$.  We choose the covariance to be proportional to the
 gain: $\tau_{ij} = \tau_0 g_{ij}$. In this example, $w_r = 0.2$,
 $w_\theta = 20^\circ$, $g_0=1$, and $\tau_0 = 0.8$. The gain matrix
 for a grid of $16\times16$ neuron populations is shown in
 \crefformat{figure}{Fig.~#2#1{(b)}#3} \cref{fig:gFunFig}.  This grid
 size requires us to solve $N=256$ self-consistent equations to
 determine the eigenvalue density. For comparison, we generated 1000
 realizations of $J$ with $N=2048$; to mitigate finite-$N$ effects
 \cite{aljadeff2015low}, we used block structured matrices with $16
 \times 16$ populations, with 8 nodes in each population. We find that
 our result closely matches the empirical distribution
 \crefformat{figure}{Fig.~#2#1{(c-e)}#3}
 (\cref{fig:gFunFig}). Increasing the grid size to $32 \times 32$ and
 $64 \times 64$ did not appreciably change the resulting eigenvalue
 distribution, indicating that the current resolution is sufficient.
 Finally, using Eqs.~(\ref{eqn:covBoundary}-\ref{eqn:gammaOutside})),
 we find that including correlations moves the rightmost eigenvalue
 from ~0.24 to ~0.41, decreasing the stability of the system.




Our results can be extended to more general correlation structures, such as correlations between arbitrary blocks or clusters. However, including more general correlations increases the number of self-consistent equations that must be solved in (\ref{eqn:blockGc}). Our results can also be extended to the case of nonzero mean as in \cite{ahmadian2015}. The diagrammatic technique can also be used to study further quantities of interest such as eigenvalue correlations \cite{janik1997non}, eigenvector correlations \cite{chalker1998eigenvector,janik1999correlations}, and linear dynamics not captured by the eigenvalues \cite{ahmadian2015}.

In conclusion, we have adapted the diagrammatic technique to study
correlated connectivity matrices that are not independently or
identically distributed, and relevant to biological circuits. The
results indicate that the presence of correlations can dramatically
influence the network stability and dynamics.
The correlation structure is determined by plasticity rules, which act
locally on connections between nodes \cite{Gilson2009,Miner_2016}. The presented
analytical framework therefore makes it possible to evaluate the
impact of local plasticity rules on global network activity.

\appendix
\section{Appendix}
\paragraph{Derivation of the boundary of the eigenvalue distribution in the absence of covariance between matrix elements} Here we first show that the eigenvalue density (\ref{eqn:polarDensity}) for $J$ with independent elements ($\tau_{ij}=0$) has support on
the disk with radius $R = \sqrt{\lambda_1(\boundaryMat)}$, where
$\lambda_1(\boundaryMat)$ is the largest eigenvalue of the matrix
$\boundaryMat_{ij} \equiv g^2_{ij}/N$.
There are two solutions to the self-consistent equations (\ref{eqn:selfConsistentNoCov}) in the limit $\ep\rightarrow 0$: a trivial solution, with all $a_i=d_i=0$, and a non-trivial solution, with all $a_i,d_i > 0$ \footnote{It is not hard to show that if just one $a_i$ or $d_i$ is zero, all are zero}. The trivial solution corresponds to the region where $\rho(r)=0$ \cite{zee1997,feinberg2006non}. Indeed, we see that when $a_i=d_i=0$, all $q_i = r^2$. Then by (\ref{eqn:selfConsistentNoCov}) $c_i = 1/z$, and therefore $\rho(r) = 0$ by (\ref{eqn:polarDensity}).

Now consider the region where $\rho \neq 0$, where all the $a_i$ and $d_i$ are nonzero. Then, combining (\ref{eqn:selfConsistentNoCov}) and (\ref{eqn:ahatdhat}) for $d_i$ in the $\ep \rightarrow 0$ limit yields
\begin{equation}
q_i d_i = \frac{1}{N} \sum_j g_{ij} d_j.
\label{eqn:qBoundary}
\end{equation}
We determine the radius $R$ of the boundary by finding where the two
solutions match. Assuming continuity of the $a_i$ and $d_i$, then as
$d_i \rightarrow 0^+$ as we approach the boundary, all the $q_i
\rightarrow R^2$. Then, in the limit, (\ref{eqn:qBoundary}) indicates
that $d$ is an eigenvector of $\boundaryMat_{ij}=g_{ij}^2/N$ with
eigenvalue $R^2$. Furthermore, since $\boundaryMat$ and $d$ have only
positive entries, $R^2$ must be the largest eigenvalue
$\lambda_1(\boundaryMat)$ of $\boundaryMat$ by the Perron-Frobenius
theorem.  Thus, the boundary of the eigenvalue distribution has radius
$R = \sqrt{\lambda_1(\boundaryMat)}$. A nearly identical argument
shows $K_{\cidx\didx}=g_{\cidx\didx}^2 f_\didx$ for the block
structured case.  This result was previously presented in
\cite{aljadeff2015eigenvalues} and \cite{aljadeff2015low}, and a
similar argument was used in \cite{ahmadian2015} for the case of
matrices with non-zero mean. However, previous analyses do not hold
when $J$ has covariant elements.

\paragraph{Boundary with covariance}
Now we show that when $\tau_{ij}\neq 0$, the boundary of the eigenvalue distribution satisfies (\ref{eqn:covBoundary}) and (\ref{eqn:gammaOutside}).
Again, we have $a_i,d_i\neq 0$ on the support of the eigenvalue distribution, and $a_i=d_i=0$ otherwise.
Plugging the trivial solution into (\ref{eqn:cCov}), the $c_i$ now satisfy 
\begin{equation}
c_i = (z-\sum_j \tau_{ij}g_{ij}g_{ji} c_j)^{-1}.
\label{eqn:cOutsideAppendix}
\end{equation}
Now, approaching the boundary from the inside as before, in the limit $d_i\rightarrow 0^+$,
\begin{equation}
d_i = \sum_j |c_i|^2 g_{ij}^2 d_j.
\end{equation}
where the $c_i$ satisfy (\ref{eqn:cOutsideAppendix}) in the limit. Since all the $d_i > 0$, this means
that $d$ is the Perron-Frobenius eigenvector of the matrix $\boundaryMat_{ij} = |c_i|^2 g_{ij}^2$ with eigenvalue $ 1$.
This means that the points $z$ on the boundary satisfy
\begin{equation}
\lambda_1(\boundaryMat) = 1
\label{eqn:maxEigAppendix}
\end{equation}
where $\lambda_1(\boundaryMat)$ is the largest modulus eigenvalue of
$\boundaryMat$. Together, (\ref{eqn:cOutsideAppendix}) and
(\ref{eqn:maxEigAppendix}) determine the points $z$ that lie on the
boundary of the eigenvalue distribution.  We have
found that these equations can be solved efficiently as follows: First
we write $z=r e^{i\theta}$ and fix $\theta$. Then, to find the $r$
satisfying (\ref{eqn:maxEigAppendix}), we use a root finding
algorithm: at each step of the root finding algorithm, we iterate
(\ref{eqn:cOutsideAppendix}) to find the $c_i(z)$.

If $g_{ij}$ is block-structured, then we have only $M$ variables $c_\cidx$, with
\begin{equation}
c_\cidx = (z-\sum_\didx \tau_{\cidx\didx} g_{\cidx\didx} g_{\didx\cidx} f_\didx c_\didx)^{-1}
\end{equation}
and
\begin{equation}
\boundaryMat_{\cidx\didx} = |c_\cidx|^2 g_{\cidx \didx}^2 f_\didx
\label{eqn:blockBoundaryMatrix}
\end{equation}
\bibliographystyle{apsrev4-1}

\end{document}